\documentclass[12pt]{article}
\usepackage{epsfig}
\usepackage{latexsym}
\usepackage{graphicx}
\usepackage{amsmath}
\usepackage{amsfonts}
\usepackage{amssymb}

\newtheorem{theorem}{Theorem}
\newtheorem{proposition}[theorem]{Proposition}

\newtheorem{corollary}[theorem]{Corollary}

\newtheorem{question}{Question}

\newcommand\efface[1]{} 

\def\cp{\,\Box\,} 
\def\sp{\,\boxtimes\,} 
\def\ddp{\times} 
\newcommand{\vv}[1]{\overrightarrow{#1}}  
\def\hom{\rightarrow} 
\def\lhom{\longrightarrow} 

\newenvironment{proof}{
\par
\noindent {\bf Proof.}\rm}{\mbox{}\hfill$\square$\par\vskip 3mm}

\begin{document}
\title{Upper oriented chromatic number of undirected graphs
and oriented colorings of product graphs}
\author{\'Eric Sopena\\
  {\small Universit\'e de Bordeaux, LaBRI UMR CNRS 5800,}\\
  {\small 351, cours de la Lib\'eration, F-33405 Talence  Cedex, France}\\
  {\small \tt eric.sopena@labri.fr}
}
\date{\today}

\maketitle

\begin{abstract}
The oriented chromatic number of an oriented graph $\vv G$ is the minimum
order of an oriented graph $\vv H$ such that $\vv G$ admits a homomorphism
to $\vv H$. The oriented chromatic number of an undirected graph $G$ is
then the greatest oriented chromatic number of its orientations.

In this paper, we introduce the new notion of the {\em upper oriented chromatic
number\,} of an undirected graph $G$, defined as the minimum order of an oriented
graph $\vv U$ such that {\em every\,} orientation $\vv G$ of $G$ admits a
homomorphism to $\vv U$. We give some properties of this parameter, derive
some general upper bounds on the ordinary and upper oriented chromatic numbers of Cartesian,
strong, direct and lexicographic products of graphs, and consider the particular case
of products of paths.
\end{abstract}

\section{Introduction}

All the graphs we consider in this paper are simple, with no loops or multiple edges. 
An {\em oriented graph\,} $\vv G=(V(\vv G), E(\vv G))$ is 
an antisymmetric digraph obtained from an undirected graph $G=(V(G),E(G))$ having the
same set of vertices, $V(G)=V(\vv G)$, by giving to each edge $\{u,v\}$ in
$E(G)$ one of its two possible orientations, $(u,v)$ or $(v,u)$. 
Such an oriented graph $\vv G$
is said to be an {\em orientation\,} of $G$.

Let $G$ and $H$ be two undirected graphs. A {\em homomorphim\,} from $G$ to
$H$ is a mapping $h:V(G)\lhom V(H)$ such that for every edge $\{u,v\}$ in $E(G)$,
$\{h(u),h(v)\}$ is an edge in $H$. We write $G\hom H$ whenever such a homomorphism
exists. Homomorphisms of oriented graphs are defined similarly, by using arcs instead
of edges.

A {\em proper $k$-coloring\,} of an undirected graph $G$ is a mapping
$c:V(G)\lhom\{1,2,\dots,k\}$ such that $c(u)\neq c(v)$ for every edge $\{u,v\}$
in $E(G)$. Such a coloring can also be viewed as a partition
of $V(G)$ into $k$ disjoint independent sets $V_1,\dots,V_k$.
The {\em chromatic number\,} $\chi(G)$ of $G$ is then the smallest $k$
for which $G$ admits a proper $k$-coloring. It is easy to see that $\chi(G)$ 
also corresponds to the smallest $k$ for which $G\hom K_k$, where $K_k$ stands
for the complete graph of order $k$.

An {\em oriented $k$-coloring\,} of an oriented graph $\vv G$ is
a partition of $V(\vv G)$ into $k$ disjoint independent sets, such that all arcs
linking any two of these sets {\em have the same direction}. Such an oriented
coloring is thus a mapping $\gamma:V(\vv G)\lhom\{1,2,\dots,k\}$ such that
$\gamma(u)\neq\gamma(v)$ for every arc $(u,v)$ in $E(\vv G)$ and
$\gamma(u)\neq\gamma(x)$ whenever there exist two arcs $(u,v)$ and $(w,x)$
in $E(\vv G)$ with $\gamma(v)=\gamma(w)$ ($v$ and $w$ are not necessarily distinct).

The {\em oriented chromatic number\,} $\chi_o(\vv G)$ of an oriented graph $\vv G$ is then defined
as the smallest $k$ for which $\vv G$ admits an oriented $k$-coloring.
As before, $\chi_o(\vv G)$ also corresponds to the smallest order of an oriented 
graph $\vv T$ for which $\vv G\hom\vv T$.
If $G$ is an {\em undirected\,} graph, the oriented chromatic number $\chi_o(G)$
of $G$ is defined as the highest oriented chromatic number of its
orientations:
$$\chi_o(G)=\max\big\{\ \chi_o(\vv G),\ \vv G\ \mbox{is an orientation of } G\ \big\}.$$

Most of the papers devoted to oriented colorings were concerned with upper bounds
on the oriented chromatic number of special classes of 
graphs~\cite{AS,EO,FRR,KSZ,NNS,O,RS,SURVEY,W}. 
In many cases,
such bounds were obtained by proving that every orientation of any graph in
a given class admits a homomorphism to some specific oriented graph (sometimes
a tournament). This observation motivates the introduction of a new parameter,
that we call the {\em upper oriented chromatic number\,} of an undirected graph $G$, defined
as the smallest order of an oriented graph $\vv T$ such that $\vv G\hom\vv T$
for {\em every\,} orientation $\vv G$ of $G$.

The aim of this paper is to initiate the study of this new parameter.
We shall give general bounds on the upper and ordinary oriented chromatic numbers
of lexicographic, strong, Cartesian and direct products of undirected and oriented 
graphs, with a particular focus on products of paths.

This paper is organised as follows. In Section~\ref{s:definitions}, we give the 
main definitions we shall use later and provide some elementary properties
of the upper oriented chromatic number. The four following sections are
respectively concerned with lexicographic, strong, Cartesian and direct
products of graphs, and contain our main results. 
Some open problems and directions for future work are discussed in Section~\ref{s:discussion}.

\section{Definitions and notation}
\label{s:definitions}

We denote by $P_k$ the path on $k$ vertices, by $C_k$ the cycle on
$k$ vertices and by $K_k$ the complete graph on $k$ vertices.
If $G$ is an undirected graph, we denote by $\{u,v\}$ an edge
linking vertices $u$ and $v$ in $G$.
If $\vv G$ is an oriented graph, we denote by $(u,v)$ an arc
directed from vertex $u$ to vertex $v$ in $\vv G$.
A {\em directed path\,} of length $k$ in an oriented graph $\vv G$ is a sequence
of vertices $x_1\dots x_k$ such that $(x_i,x_{i+1})\in E(\vv G)$ for
every $i$, $1\le i<k$.
We shall denote by $\vv{DP_k}$ the particular orientation of the
path $P_k$ corresponding to the directed path of length $k$, while $\vv{P_k}$
will denote any orientation of $P_k$.
For any two sets $A$ and $B$,
the Cartesian product $A\times B$ denotes the set
$\big\{\ [a,b],\ a\in A,\ b\in B\ \big\}$.

An {\em oriented clique} $\vv H$ is an oriented graph in which every pair of
vertices is joined by a directed path of length 1 or 2. From the definition
of oriented colorings, it follows that $\chi_o(\vv H)=|V(\vv H)|$, whenever
$\vv H$ is an oriented clique.

We define the {\em upper oriented chromatic number\,} of an undirected graph $G$,
denoted $\chi_o^+(G)$, as the smallest order of an oriented graph $\vv T$
such that $\vv G\hom \vv T$ for every orientation $\vv G$ of $G$.
The property of having upper oriented chromatic number at most $k$
is {\em hereditary}, that is $\chi_o^+(H)\le\chi_o^+(G)$ for every
subgraph $H$ of $G$.

From this definition, we clearly have:

\begin{proposition}
For every
undirected graph $G$, $\chi_o(G)\le\chi_o^+(G)$. 
\end{proposition}

Consider for instance the cycle $C_3$ on three vertices.
We  have $\chi_o(C_3)=3$ since $\chi_o(G)\le|V(G)|$ for every undirected
graph $G$ and any two vertices in any orientation $\vv{C_3}$ of $C_3$ are linked
by a directed path of length 1 or 2. However, $\chi_o^+(C_3)=4$ since the smallest
oriented graph contained both a directed 3-cycle and a transitive 3-cycle as subgraphs
has four vertices.
 
More generally, for every $n\ge 3$, $\chi_o(K_n)=n$ and $\chi_o^+(K_n)=\varepsilon(n)$,
where $\varepsilon(n)$ stands for the minimum size of an {\em $n$-universal\,} tournament,
that is a tournament containing every tournament of order $n$ as a subgraph. 
The following result is proved in~\cite[Chapter~17]{MOON}:

\begin{theorem}
For every $n\ge 1$,
$$2^\frac{n-1}{2}\le\varepsilon(n)\le 
\left\{ 
\begin{array}{ll}
n\,2^\frac{n-1}{2} & \mbox{if $n$ is odd,} \\
\frac{3}{2\sqrt{2}}\,n\,2^\frac{n-1}{2} &  \mbox{if $n$ even.}
\end{array}
\right.
$$
\label{th:moon}
\end{theorem}

Hence, the difference between
$\chi_o(G)$ and $\chi_o^+(G)$ can be arbitrarily large.

For every graph $G$, let $\omega(G)$ denote the {\em clique number\,}
of $G$, that is the maximum order of a complete subgraph of $G$.
From Theorem~\ref{th:moon}, we get the following general lower and upper bounds on
the upper oriented chromatic number of any graph:

\begin{corollary}
For every graph $G$ of order $n$ and clique number $\omega(G)$,
$$\chi_o^+(G)\ge\chi_o^+(K_{\omega(G)})\ge 2^\frac{\omega(G)-1}{2}$$
and
$$\chi_o^+(G)\le\chi_o^+(K_n)\le
\left\{ 
\begin{array}{ll}
n\,2^\frac{n-1}{2} & \mbox{if $n$ is odd,} \\
\frac{3}{2\sqrt{2}}\,n\,2^\frac{n-1}{2} &  \mbox{if $n$ even.}
\end{array}
\right.
$$
\label{cor:moon}
\end{corollary}

For complete bipartite graphs, we have the following:

\begin{theorem}
For every $m,n$ with $m\le n$,
$$\chi_o(K_{m,n})=m+\min\{n,2^m\}$$
and $$\chi_o^+(K_{m,n})\le m+2^m.$$
\label{th:bipartite}
\end{theorem}

\begin{proof}
Let $\{x_1,\dots,x_m\}$ and $\{y_1,\dots,y_n\}$ denote
the two maximal independent sets of $K_{m,n}$.
We first consider $\chi_o^+(K_{m,n})$. 
Let $\vv T$ be the oriented graph defined by
$$V(\vv T)=\{a_1,\dots,a_m\}\cup\left\{b_S,\ S\subseteq\{1,2,\dots,m\}\right\}$$
and $(a_i,b_S)\in E(\vv T)$ if and only if $i\in S$.
Let now $\vv K_{m,n}$ be any orientation of $K_{m,n}$.
For every $j$, $1\le j\le n$, let $N^-(y_j)=\{x_i,\ (x_i,y_j)\in E(\vv K_{m,n})\}$.
Clearly, the mapping $\varphi:V(\vv K_{m,n})\lhom V(\vv T)$
defined by $\varphi(x_i)=a_i$ for every $i$, $1\le i\le m$,
and $\varphi(y_j)=b_{N^-(y_j)}$ for every $j$, $1\le j\le n$,
is a homomorphism and, therefore,
$\chi_o^+(K_{m,n})\le m+2^m$.

We now consider $\chi_o^+(K_{m,n})$.
Suppose first that $n\ge 2^m$ and let $S_1,\dots,S_{2^m}$ denote
the $2^m$ subsets of $\{1,2,\dots,m\}$.
Let now  $\vv K_{m,n}$ be the orientation of $K_{m,n}$ defined by
$(x_i,y_j)\in E(\vv K_{m,n})$ if and only if $j\le 2^m$ and $i\in S_j$,
and let $\vv X$ denote the subgraph of $\vv K_{m,n}$ induced
by the set of vertices $\{x_1,\dots,x_m\}\cup\{y_1,\dots,y_{2^m}\}$.
The subgraph $\vv X$ is clearly an oriented clique
and, therefore, $\chi_o(K_{m,n})\ge\chi_o(\vv X)=m+2^m$. On the other hand,
$\chi_o(K_{m,n})\le\chi_o^+(K_{m,n})\le m+2^m$.

If $n<2^m$, we get $\chi_o(K_{m,n})\ge|V(K_{m,n})|=m+n$ by considering the
orientation of $K_{m,n}$ corresponding to the subgraph of the above defined
oriented graph $\vv X$ induced by the set of vertices 
$\{x_1,\dots,x_m\}\cup\{y_1,\dots,y_n\}$. Since 
$\chi_o(K_{m,n})\le|V(K_{m,n})|$, the result follows.
\end{proof}

As said before, many upper bounds on the oriented chromatic number
of classes of undirected graphs have been obtained by providing some
oriented graph $\vv T$ such that every orientation of every undirected graph
in the class admits a homomorphism to $\vv T$. Therefore, every such upper
bound also holds for the upper oriented chromatic number of the same graphs.
The following theorem summarizes some of these results. 
Let us recall that a graph is {\em outerplanar\,} if it has a planar drawing with all
vertices lying on the outer face. 
A graph is {\em 2-outerplanar\,} whenever it admits a planar drawing such that
deleting all the vertices of the outer face produces an outerplanar graph.
The
{\em acyclic chromatic number\,} of an undirected graph $G$ is the smallest
number of colors needed in an {\em acyclic coloring\,} of $G$, that is
a proper coloring of $G$ in which every cycle uses at least three colors.

\begin{theorem} Let $G$ be an undirected graph. We then have:
\begin{enumerate}
\setlength{\itemsep}{1pt}
\setlength{\parskip}{0pt}
\setlength{\parsep}{0pt}
\item If $G$ is a forest with at least three vertices, then $\chi_o^+(G)=3$~{\rm\cite{SURVEY}},
\item If $G=C_k$ is a cycle on $k\ge 3$ vertices, then $\chi_o^+(G)=4$ when $k\neq 5$,
       and $\chi_o^+(G)=5$ when $k=5$~{\rm\cite{SURVEY}},
\item If $\chi_a(G)\le a$, then $\chi_o^+(G)\le a2^{a-1}$~{\rm\cite{RS}}, 
and this bound is tight for $a\ge 3$~{\rm\cite{O}},
\item If $G$ is an outerplanar graph, then $\chi_o^+(G)\le 7$ and this bound is tight~{\rm\cite{SURVEY}},
\item If $G$ is a 2-outerplanar graph, then $\chi_o^+(G)\le 67$~{\rm\cite{EO}},
\item If $G$ is a planar graph, then $\chi_o^+(G)\le 80$~{\rm\cite{RS}},
\item If $G$ is a triangle-free planar graph, then $\chi_o^+(G)\le 59$~{\rm\cite{O2}},
\item If $G$ has maximum degree $\Delta(G)=k$, then $\chi_o^+(G)\le 2k^22^k$~{\rm\cite{KSZ}}.
\end{enumerate}
\end{theorem} 

Let $G$ be an undirected graph. The {\em square\,} of $G$ is the graph
$G^2$ defined by $V(G^2)=V(G)$ and
$E(G^2)=\left\{\{u,v\},\ 1\le d_G(u,v)\le 2\right\}$, where $d_G(u,v)$ denotes the distance
between vertices $u$ and $v$ in $G$.
The following theorem provides an upper bound on the upper oriented
chromatic number of an undirected graph depending on the chromatic
number of its square:

\begin{proposition}
For every undirected graph $G$ with $k=\chi(G^2)$, 
$\chi_o^+(G)\le 2^k-1$.
\label{prop:upper-square}
\end{proposition}

\begin{proof}
Let $\sigma$ be a proper $k$-coloring of $G^2$ and $S$ be the
set of $2^k-1$ elements defined as
$$S=\left\{[a,b_1,\dots,b_{a-1}],\ 1\le a\le k,\ b_i\in\{0,1\},\ 1\le i\le a-1\right\}.$$
Let now $\vv G$ be any orientation of $G$.
We define a mapping $\varphi$ from $V(G)$ to $S$
as follows. For every vertex $u\in V(G)$, we set
$\varphi(u)=[a(u),b_1(u),\dots,b_{a(u)-1}(u)]$ where:
\begin{enumerate}
\setlength{\itemsep}{1pt}
\setlength{\parskip}{0pt}
\setlength{\parsep}{0pt}
\item $a(u)=\sigma(u)$,
\item for every $i$, $1\le i\le a(u)-1$, if $u$ has a neighbor $v$
with $\sigma(v)=i$ such that $(v,u)$ is an arc in $E(\vv G)$ then
$a_i(u)=1$, otherwise $a_i(u)=0$. (Since $\sigma$ is a proper coloring
of $G^2$, if such a vertex $v$ exists it must be unique.)
\end{enumerate}
We claim that $\varphi$ is an oriented coloring of $\vv G$.
Observe first that if $u$ and $v$ are adjacent vertices in $\vv G$,
then $\varphi(u)\neq\varphi(v)$ since $\sigma(u)\neq\sigma(v)$.
Suppose now that there exist two arcs $(u,v)$ and $(w,x)$ in $E(\vv G)$
such that $\varphi(u)=\varphi(x)$ and $\varphi(v)=\varphi(w)$.
If $u=x$, then $\sigma(v)\neq\sigma(w)$ since $\sigma$ is a proper coloring
of $G^2$, a contradiction.
Otherwise, we may assume without loss of generality
that $i=\sigma(u)=\sigma(x)<\sigma(v)=\sigma(w)$.
Since $(u,v),(w,x)\in E(\vv G)$, we get $b_i(v)=1$ and $b_i(w)=0$, again a contradiction.

The mapping $\varphi$ is thus an oriented coloring of $\vv G$ using at most
$2^k-1$ colors and the result follows.
\end{proof}

Since $\chi(G^2)\ge\Delta(G)+1$ for every graph $G$,
Proposition~\ref{prop:upper-square} does not give an interesting general bound
for classes of graphs with unbounded degree.

Let $G$ and $H$ be two undirected graphs.
The {\em Cartesian product\,} of $G$ and $H$ is the undirected
graph $G\cp H$ defined by $V(G\cp H)=V(G)\times V(H)$ and
$\{[u,v],[u',v']\}$ is an edge in $E(G\cp H)$ if and only if
either $u=u'$ and $\{v,v'\}\in E(H)$ or $v=v'$ and $\{u,u'\}\in E(G)$.

The {\em strong product\,} of $G$ and $H$ is the undirected
graph $G\sp H$ defined by $V(G\sp H)=V(G)\times V(H)$ and
$\{[u,v],[u',v']\}$ is an edge in $E(G\sp H)$ if and only if
either $u=u'$ and $\{v,v'\}\in E(H)$ or $v=v'$ and $\{u,u'\}\in E(G)$
or $\{u,u'\}\in E(G)$ and $\{v,v'\}\in E(H)$.

The {\em direct product\,} of $G$ and $H$ is the undirected
graph $G\ddp H$ defined by $V(G\ddp H)=V(G)\times V(H)$ and
$\{[u,v],[u',v']\}$ is an edge in $E(G\ddp H)$ if and only if 
$\{u,u'\}\in E(G)$ and $\{v,v'\}\in E(H)$.

The {\em lexicographic product\,} of $G$ and $H$ is the undirected
graph $G[H]$ defined by $V(G[H])=V(G)\times V(H)$ and
$\{[u,v],[u',v']\}$ is an edge in $E(G[H])$ if and only if 
either $\{u,u'\}\in E(G)$ or $u=u'$ and $\{v,v'\}\in E(H)$.

Cartesian, strong, direct and lexicographic products of {\em oriented\,} graphs
are defined similarly, by replacing edges by arcs in the above
definitions.

It is not difficult to see that the Cartesian, strong and direct products are symmetric
operations while $G[H]$ and $H[G]$ are generally not isomorphic graphs (this justifies
our notation for the lexicographic product). However, these four products are associative. 
Moreover, $G\cp H\subseteq G\sp H$, $G\ddp H\subseteq G\sp H$ and
$G\sp H\subseteq G[H]$ for every 
undirected or oriented graphs 
$G$ and $H$. 
Therefore, every upper bound on the upper (or ordinary) oriented chromatic number
of $G[H]$ (resp. $G\sp H$) holds for $G\sp H$ (resp. $G\cp H$ and $G\ddp H$).

Following the reference book of Imrich and Klav\u zar~\cite{IK}, we shall denote
by $G_v$ (resp. $H_u$) the {\em $v$-layer\,} (resp. the {\em $u$-layer}) of
$G\sp H$, $G\cp H$ or $G\ddp H$, that is the subgraph
induced by $V(G)\times\{v\}$ (resp. $\{u\}\times V(H)$), for every $v\in V(H)$
(resp. $u\in V(G)$).

The study of the oriented chromatic number of Cartesian and strong products
of graphs has been recently initiated by Natarajan, Narayanan and Subramanian~\cite{NNS}.
In particular, they proved that for every undirected graph $G$,
$\chi_o(G\cp P_k)\le (2k-1)\chi_o(G)$ and $\chi_o(G\cp C_k)\le 2k\chi_o(G)$ for every
$k\ge 3$. We shall improve these two bounds in Section~\ref{s:cartesian}.

\section{Lexicographic products}
\label{s:lexicographic}

Concerning the oriented chromatic number of the lexicographic product of oriented
graphs, we have the following:

\begin{theorem}
If $\vv{G}$, $\vv{H}$, $\vv{T}$ and $\vv{U}$ are oriented graphs such that
$\vv{G}\hom\vv{T}$ and $\vv{H}\hom\vv{U}$, then
$\vv{G}[\vv{H}]\hom\vv{T}[\vv{U}]$.
Therefore, for every oriented graphs 
$\vv{G}$ and $\vv{H}$,
$$\chi_o(\vv{G}[\vv{H}])\le \chi_o(\vv{G})\chi_o(\vv{H}).$$ 
\label{th:lexicographic-oriented}
\end{theorem}

\begin{proof}
Let $\alpha:\vv{G}\hom\vv{T}$
and $\beta:\vv{H}\hom\vv{U}$ be two homomorphisms.
Let now  $\varphi:V(\vv{G}[\vv{H}])\lhom V(\vv{T}[\vv{U}])$ be the 
mapping defined by 
$\varphi([u,v])=[\alpha(u),\beta(v)]$ for every vertex $[u,v]$ in $V(\vv{G}[\vv{H}])$.

We claim that $\varphi$ is a homomorphism. To see this, let $([u,v],[u',v'])$ be an
arc in $\vv{G}[\vv{H}]$.
We then have $\varphi([u,v])=[\alpha(u),\beta(v)]$ and 
$\varphi([u',v'])=[\alpha(u'),\beta(v')]$.
If $(u,u')\in E(\vv{G})$, then $(\alpha(u),\alpha(u'))\in E(\vv T)$.
If $u=u'$ and $(v,v')\in E(\vv{H})$, then $\alpha(u)=\alpha(u')$ and 
$(\beta(v),\beta(v'))\in E(\vv{U})$.
Therefore, every arc in $\vv{G}[\vv{H}]$ 
is mapped
to an arc in $\vv{T}[\vv{U}]$ and $\varphi$ is a homomorphism.

The inequality $\chi_o(\vv{G}[\vv{H}])\le \chi_o(\vv{G})\chi_o(\vv{H})$
directly follows from the definition of the
oriented chromatic number.
\efface{To see that this bound is tight, it suffices to consider the lexicographic product
of any two tournaments $\vv T_1$ and $\vv T_2$. In that case, $\vv T_1[\vv T_2]$ is also
a tournament and 
$\chi_o(\vv T_1[\vv T_2])=|V(\vv T_1[\vv T_2])|=|V(\vv T_1)||V(\vv T_2)|=\chi_o(\vv T_1)\chi_o(\vv T_2)$.
(More generally, this equality holds for every two {\em oriented cliques}, that is oriented
graphs whose any two vertices are linked by a directed path of length 1 or 2, since
any such lexicographic product is also an oriented clique.)
} 
\end{proof}

For the lexicographic product of directed paths, we have the following:

\begin{theorem}
For every $k,\ell\ge 3$,
$\chi_o(\vv{DP_k}[\vv{DP_\ell}])=9$. Therefore, the bound given in Theorem~\ref{th:lexicographic-oriented}
is tight.
\label{th:lexicographic-oriented-paths}
\end{theorem}

\begin{proof}
Since $\chi_o(\vv P)\le 3$ for every oriented path $\vv P$, we have
$\chi_o(\vv{DP_k}[\vv{DP_\ell}])\le 9$ by Theorem~\ref{th:lexicographic-oriented}.
Since any two vertices in $\vv{DP}_3[\vv{DP}_3]$ are linked by a directed path of length 1 or 2,
we have $\chi_o(\vv{DP_k}[\vv{DP_\ell}])\ge\chi_o(\vv{DP}_3[\vv{DP}_3])=9$ for every $k,\ell\ge 3$.
\end{proof}

The following result provides a general upper bound on the upper oriented
chromatic number of lexicographic products of undirected graphs.

\begin{theorem}
Let $G$ and $H$ be two undirected graphs with $k=\chi(G^2)$ and $n=|V(H)|$. We then have
$$\chi_o^+(G[H])\le k(n+2^n)^{k-1}\chi_o^+(H).$$
\label{th:lexico-undirected}
\end{theorem}

\begin{proof}
Let $\alpha$ be a proper $k$-coloring of $G^2$ and $\ell=\chi_o^+(H)$.
Let $\vv U$ be an oriented graph of order $\ell$ such that $\vv H\hom\vv U$ for
every orientation $\vv H$ of $H$
and $\vv T$ be the oriented graph of order $n+2^n$ such that $\vv K_{n,n}\hom\vv T$ for every
orientation $\vv K_{n,n}$ of $K_{n,n}$, as defined in the proof
of Theorem~\ref{th:bipartite} (we consider here the case $m=n$).
We define the digraph $\vv W$ by
$$V(\vv W)=\big\{\ [a,b,c_1,\dots,c_k],\ 
1\le a\le k,\ 1\le b\le \ell,$$
$$c_a=0,\ 1\le c_i\le n+2^n,\ 1\le i\le k,\ i\neq a\ \big\}$$
and $([a,b,c_1,\dots,c_k],[a',b',c'_1,\dots,c'_k])$ is an arc in $E(\vv W)$
if and only if either $a=a'$ and $(b,b')\in E(\vv U)$,
or $a\neq a'$ and $(c_{a'},c'_a)\in E(\vv T)$.
The digraph $\vv W$ is clearly an oriented graph of order $k\ell(n+2^n)^{k-1}$.

Let now $\vv{G[H]}$ be any orientation of $G[H]$,
$\vv{H_u}$ be the oriented copy of $H$ induced by the set of vertices $\{u\}\times V(H)$ and 
$\lambda_u:\vv{H_u}\hom\vv U$ be a homomorphism. 
If $\{u,u'\}$ is an edge in $G$, let $\mu_{u,u'}$ be a homomorphism of the oriented subgraph
of $\vv{G[H]}$ induced by the set of vertices $\{u,u'\}\times V(H)$ to $\vv T$.
We shall now construct a mapping
$\varphi$ from $V(\vv{G[H]})$ to $V(\vv W)$, and prove that this mapping is a homomorphism,
which will give the desired result.

Let $[u,v]$ be any vertex in $V(\vv{G[H]})$ and $\varphi([u,v])=[a,b,c_1,\dots,c_k]$
be defined as follows:
\begin{enumerate}
\renewcommand{\labelenumi}{(\roman{enumi})}
\setlength{\itemsep}{1pt}
\setlength{\parskip}{0pt}
\setlength{\parsep}{0pt}
\item $a=\alpha(u)$,
\item $b=\lambda_u(v)$,
\item if there is a neighbor $u'$ of $u$ in $G$ with $\alpha(u')=a'$, then
$c_{a'}=\mu_{u,u'}([u,v])$, otherwise $c_{a'}=0$.
\end{enumerate}
Note that if such a neighbor $u'$ of $u$ exists in item (iii), then it must be unique since
$\alpha$ is a proper coloring of $G^2$.

Let now $([u,v],[u',v'])$ be any arc in $E(\vv{G[H]})$,
$\varphi([u,v])=[a,b,c_1,\dots,c_k]$ and
$\varphi([u',v'])=[a',b',c'_1,\dots,c'_k]$.
If $u=u'$, then $a=\alpha(u)=\alpha(u')=a'$ and $(b,b')=(\lambda_u(v),\lambda_u(v'))\in E(\vv U)$.
If $u\neq u'$, then $\{u,u'\}\in E(G)$, $c_{a'}=\mu_{u,u'}([u,v])$, $c'_a=\mu_{u,u'}([u',v'])$ and,
therefore, $(c_{a'},c'_a)\in E(\vv T)$.
Every arc of $\vv{G[H]}$ is thus mapped to an arc of $\vv W$ and $\varphi$ is a homomorphism.
\end{proof}

Since $\chi(P^2)\le 3$ and $\chi_o^+(P)\le 3$ for every path $P$,
Theorem~\ref{th:lexico-undirected} implies
that for every $k,\ell\ge 3$, $\chi_o^+(P_k[P_\ell])\le 9(\ell+2^\ell)^2$.

\section{Strong products}
\label{s:strong}

Since $\vv G\sp\vv H\subseteq\vv G[\vv H]$ for every two oriented
graphs $\vv G$ and $\vv H$, Theorem~\ref{th:lexicographic-oriented}
implies the following:

\begin{corollary}
For every oriented graphs $\vv{G}$ and $\vv{H}$,
$$\chi_o(\vv{G}\sp\vv{H})\le\chi_o(\vv G)\chi_o(\vv H).$$
\label{cor:strong-oriented}
\end{corollary}

This result can be strengthtened as follows:

\begin{theorem}
If $\vv{G}$, $\vv{H}$, $\vv{T}$ and $\vv{U}$ are oriented graphs such that
$\vv{G}\hom\vv{T}$ and $\vv{H}\hom\vv{U}$, then
$\vv{G}\sp\vv{H}\hom\vv{T}\sp\vv{U}$.
\label{th:strong-oriented-hom}
\end{theorem}

\begin{proof}
Let $\alpha:\vv{G}\hom\vv{T}$
and $\beta:\vv{H}\hom\vv{U}$ be two homomorphisms.
Let now  $\varphi:V(\vv{G}\sp\vv{H})\lhom V(\vv{T}\sp\vv{U})$ be the 
mapping defined by 
$\varphi([u,v])=[\alpha(u),\beta(v)]$ for every vertex $[u,v]$ in $V(\vv{G}\sp\vv{H})$.

We claim that $\varphi$ is a homomorphism. To see this, let $([u,v],[u',v'])$ be an
arc in $\vv{G}\sp\vv{H}$.
We then have $\varphi([u,v])=[\alpha(u),\beta(v)]$ and 
$\varphi([u',v'])=[\alpha(u'),\beta(v')]$.
If $u=u'$ and $(v,v')\in E(\vv{H})$, then $\alpha(u)=\alpha(u')$ and 
$(\beta(v),\beta(v'))\in E(\vv{U})$.
Similarly, if $v=v'$ and $(u,u')\in E(\vv{G})$, then $\beta(v)=\beta(v')$ and 
$(\alpha(u),\alpha(u'))\in E(\vv{T})$. 
Finally, if $(u,u')\in E(\vv{G})$ and $(v,v')\in E(\vv{H})$, then
$(\alpha(u),\alpha(u'))\in E(\vv{T})$ and $(\beta(v),\beta(v'))\in E(\vv{U})$.
Therefore, every arc in $\vv{G}\sp\vv{H}$ 
is mapped
to an arc in $\vv{T}\sp\vv{U}$ and $\varphi$ is a homomorphism.
\end{proof}

Since $\chi_o(\vv{P})=3$ for every oriented path $\vv{P}$, 
Corollary~\ref{cor:strong-oriented}
gives that the oriented chromatic number of the strong product of any
two oriented paths is at most 9.
We can decrease this bound to 7 for directed paths and show that this
new bound is tight:

\begin{theorem}
For every $k,\ell\ge 3$, $\chi_o(\vv{DP_k}\sp\vv{DP_\ell})=7$.
\label{th:strong-directed-paths}
\end{theorem}

\begin{proof}
Let $\vv{DP_k}=x_1\dots x_k$ and $\vv{DP_\ell}=y_1\dots y_{\ell}$.
All arcs in $E(\vv{DP_k}\sp\vv{DP_\ell})$ are either of the form
$([x_i,y_j],[x_{i+1}y_j])$,
or $([x_i,y_j],[x_iy_{j+1}])$,
or $([x_i,y_j],[x_{i+1}y_{j+1}])$.
Let $\vv T_7=\vv T(7;1,2,3)$ be the circulant tournament defined
by $V(\vv T_7)=\{0,1,\dots,6\}$ and $(i,j)\in E(\vv T_7)$ if and only
if $(j-i)\mod 7\in\{1,2,3\}$. We will show that $\vv{DP_k}\sp\vv{DP_\ell}$
admits a homomorphism to $\vv T_7$, which proves $\chi_o(\vv{DP_k}\sp\vv{DP_\ell})\le 7$.

Let $\varphi:V(\vv{DP_k}\sp\vv{DP_\ell})\lhom V(\vv T_7)$ be the mapping defined by
$\varphi([x_i,y_j])=2j+i\pmod 7$, for every $i,j$, $1\le i\le k$, $1\le j\le\ell$.
For every arc $(u,v)$ in $\vv{DP_k}\sp\vv{DP_\ell}$, we claim that
$\varphi(v)-\varphi(u)\in\{1,2,3\}$.
If $(u,v)$ is of the form $([x_i,y_j],[x_{i+1}y_j])$, then
$\varphi(v)-\varphi(u)=2j+i+1-2j-i=1$.
If $(u,v)$ is of the form $([x_i,y_j],[x_iy_{j+1}])$, then
$\varphi(v)-\varphi(u)=2(j+1)+i-2j-i=2$.
Finally, if $(u,v)$ is of the form $([x_i,y_j],[x_{i+1}y_{j+1}])$, then
$\varphi(v)-\varphi(u))=2(j+1)+i+1-2j-i=3$.
Every arc of $\vv{DP_k}\sp\vv{DP_\ell}$ is thus mapped to an arc of $\vv T_7$ 
and $\varphi$ is a homomorphism.

To see that this bound is tight, it is enough to observe that 
any two vertices in the subgraph
$\vv X$ of $\vv{DP_k}\sp\vv{DP_\ell}$ induced by the set of vertices
$$\{\ [x_1,y_1],\ [x_2,y_1],\ [x_2,y_2],\ [x_2,y_3],\ [x_3,y_1],\ [x_3,y_2],\ [x_3,y_3]\ \}$$
are linked by a directed path
of length 1 or 2. Therefore, $\chi_o(\vv{DP_k}\sp\vv{DP_\ell})\ge\chi_o(\vv X)=7$
and the result follows.
\end{proof}

The following result provides a general upper bound on the upper oriented
chromatic number of strong products of undirected graphs.

\begin{theorem}
For every undirected graphs $G$ and $H$, 
$$\chi_o^+(G\sp H)\le(2^{\chi(H^2)}-1)\chi(G)\chi_o^+(G)\chi_o^+(H).$$
\end{theorem}

\begin{proof}
Let $k=\chi(H^2)$, $\ell=\chi(G)$, $m=\chi_o^+(G)$ and $n=\chi_o^+(H)$.
Moreover, let $\vv T$ be an oriented graph of order $m$ such that $\vv G\hom\vv T$ for every
orientation $\vv G$ of $G$ and
$\vv U$ be an oriented graph of order $n$ such that $\vv H\hom\vv U$ for every
orientation $\vv H$ of $H$.

Let now $\vv{W}$ be the digraph defined by
$$V(\vv{W})=\left\{[\alpha,\beta,\mu,\lambda,c_1, \dots, c_{\beta-1}],\ \alpha\in\{1,2,\dots,\ell\},\ 
\beta\in\{1,2,\dots,k\},\right.$$
$$\left. \mu\in\{1,2,\dots,m\},\ \lambda\in\{1,2,\dots,n\},\  c_i\in\{0,1\},\ 1\le i\le \beta-1\right\}$$
and $([\alpha,\beta,\mu,\lambda,c_1,  \dots, c_k],[\alpha',\beta',\mu',\lambda',c'_1, \dots, c'_k])$ 
is an arc in $E(\vv{W})$ if and only if one
of the following holds:
\begin{enumerate}
\renewcommand{\labelenumi}{(\roman{enumi})}
\setlength{\itemsep}{1pt}
\setlength{\parskip}{0pt}
\setlength{\parsep}{0pt}
\item $\alpha=\alpha'$ and $(\lambda,\lambda')\in E(\vv U)$,
\item $\alpha\neq\alpha'$, $\beta=\beta'$ and $(\mu,\mu')\in E(\vv T)$,
\item  $\alpha\neq\alpha'$, $\beta<\beta'$, and $c'_\beta=1$,
\item $\alpha\neq\alpha'$, $\beta>\beta'$, and $c_{\beta'}=0$.
\end{enumerate}

The graph $\vv W$ is clearly an oriented graph (with no opposite arcs)
of order $(2^k-1)\ell mn$.

Let now $\vv{G\sp H}$ be any orientation of $G\sp H$,
$\gamma$ be a proper $\ell$-coloring of $G$,
and $h$ be a proper $k$-coloring of $H^2$.
For every vertex $u\in V(G)$, let 
$\mu_u:\vv{H_u}\hom\vv U$ be a homomorphism.
Similarly, for every vertex $v\in V(H)$, let  
$\lambda_v:\vv{G_v} \hom\vv T$ be a homomorphism.

We shall now construct a mapping $\varphi$ from $V(\vv{G\sp H})$ to $V(\vv W)$, 
and prove that this mapping is a homomorphism, which
will give the desired result.

Let $[u,v]$ be any vertex in $V(\vv{G\sp H})$ and
$\varphi([u,v])=[\alpha,\beta,\mu,\lambda,c_1,\dots,c_k]$ be defined
as follows:
\begin{enumerate}
\renewcommand{\labelenumi}{(\roman{enumi})}
\setlength{\itemsep}{1pt}
\setlength{\parskip}{0pt}
\setlength{\parsep}{0pt}
\item $\alpha = \gamma(u)$,
\item $\beta=h(v)$,
\item $\mu=\mu_v(u)$,
\item $\lambda=\lambda_u(v)$,
\item if there is an arc $([w,x],[u,v])$ in $E(\vv{G\sp H})$
such that $u\neq w$ and $h(x)<h(v)$, then $c_{h(x)}=1$,
\item if there is an arc $([u,v],[w,x])$ in $E(\vv{G\sp H})$
such that $u\neq w$ and  $h(x)<h(v)$, then $c_{h(x)}=0$,
\item every $c_i$ that has not been set in (v) or (vi) is set to $0$.
\end{enumerate}
Note that in items (v) and (vi) above, if such an arc exists, then it must be unique
since $h$ is a proper coloring of $H^2$.

Let now $([u,v],[u',v'])$ be any arc in $E(\vv{G\sp H})$,
$\varphi([u,v])=[\alpha,\beta,\mu,\lambda,c_1,\dots,c_k]$,
and $\varphi([u',v'])=[\alpha',\beta',\mu',\lambda',c'_1,\dots,c'_k]$.
If $u=u'$ then $\alpha=\alpha'$ and $(\lambda,\lambda')=(\lambda_u(v),\lambda_u(v'))\in E(\vv U)$.
Similarly, if $v=v'$ then $\beta=\beta'$ and $(\mu,\mu')=(\mu_v(u),\mu_v(u'))\in E(\vv T)$.
Now, if $u\neq u'$ and $v\neq v'$, then
$\alpha\neq\alpha'$, $\beta\neq\beta'$, and 
either $\beta<\beta'$, in which case $c'_\beta=1$,
or $\beta>\beta'$, in which case $c_{\beta'}=0$.
Every arc of $\vv{G\sp H}$ is thus mapped to an arc of $\vv W$ 
and $\varphi$ is a homomorphism.
\end{proof}

Since $\chi(P)=2$ and $\chi(P^2)=\chi_o^+(P)=3$ for every path $P$, we get the following:

\begin{corollary}
For every $k,\ell\ge 3$, $\chi_o^+(P_k\sp P_\ell)\le (2^3-1).2.3.3 = 126$.
\end{corollary}

For $k=2$ and $k=3$,
Natarajan, Narayanan and Subramanian~\cite{NNS} obtained better bounds
by (implicitely) proving
that $\chi_o^+(P_2\sp P_\ell)\le 11$ and $\chi_o^+(P_3\sp P_\ell)\le 67$,
for every $\ell\ge 3$.

\section{Cartesian products}
\label{s:cartesian}

Since $\vv G\cp\vv H\subseteq\vv G[\vv H]$ for every two oriented
graphs $\vv G$ and $\vv H$, Theorem~\ref{th:lexicographic-oriented}
also implies the following:

\begin{corollary}
For every oriented graphs $\vv{G}$ and $\vv{H}$,
$$\chi_o(\vv{G}\cp\vv{H})\le\chi_o(\vv G)\chi_o(\vv H).$$
\label{cor:cartesian-oriented}
\end{corollary}

As before, this result can be strengthtened as follows:

\begin{theorem}
If $\vv{G}$, $\vv{H}$, $\vv{T}$ and $\vv{U}$ are oriented graphs such that
$\vv{G}\hom\vv{T}$ and $\vv{H}\hom\vv{U}$, then
$\vv{G}\cp\vv{H}\hom\vv{T}\cp\vv{U}$.
\label{th:cartesian-oriented-hom}
\end{theorem}

\begin{proof}
The proof is similar to the proof of Theorem~\ref{th:strong-oriented-hom}.
Let $\alpha:\vv{G}\hom\vv{T}$
and $\beta:\vv{H}\hom\vv{U}$ be two homomorphisms.
Let now  $\varphi:V(\vv{G}\cp\vv{H})\lhom V(\vv{T}\cp\vv{U})$ be the 
mapping defined by 
$\varphi([u,v])=[\alpha(u),\beta(v)]$ for every vertex $[u,v]$ in $V(\vv{G}\cp\vv{H})$.

We claim that $\varphi$ is a homomorphism. To see this, let $([u,v],[u',v'])$ be an
arc in $\vv{G}\cp\vv{H}$.
We then have $\varphi([u,v])=[\alpha(u),\beta(v)]$ and 
$\varphi([u',v'])=[\alpha(u'),\beta(v')]$.
If $u=u'$ and $(v,v')\in E(\vv{H})$, then $\alpha(u)=\alpha(u')$ and 
$(\beta(v),\beta(v'))\in E(\vv{U})$.
Similarly, if $v=v'$ and $(u,u')\in E(\vv{G})$, then $\beta(v)=\beta(v')$ and 
$(\alpha(u),\alpha(u'))\in E(\vv{T})$. Therefore, every arc in $\vv{G}\cp\vv{H}$ 
is mapped
to an arc in $\vv{T}\cp\vv{U}$ and $\varphi$ is a homomorphism.
\end{proof}

Corollary~\ref{cor:cartesian-oriented} implies that the oriented chromatic number
of the Cartesian product of two oriented paths is at most 9.
From Theorem~\ref{th:strong-directed-paths}, we get that the oriented chromatic
number of the Cartesian product of any two {\em directed\,} paths is at most 7. These
two bounds
can be improved as follows (this result also follows from a result of Natarajan
{\em et al.}~\cite{NNS}):

\begin{theorem}
For every oriented paths $\vv{P_k}$ and $\vv{P_\ell}$, $k,\ell\ge 1$, $\chi_o(\vv{P_k}\cp\vv{P_\ell})\le 3$.
\label{th:cartesian-directed-paths}
\end{theorem}

\begin{proof}
Let $\vv{P_k}=x_1\dots x_k$ and $\vv{P_\ell}=y_1\dots y_{\ell}$.
Let $\vv{C_3}$ be the directed cycle on three vertices given by
$V(\vv{C_3})=\{0,1,2\}$ and $E(\vv{C_3})=\{(0,1),(1,2),(2,0)\}$.
We inductively define a mapping $\varphi:V(\vv{P_k}\cp\vv{P_\ell})\lhom V(\vv{C_3})$ as follows:
\begin{enumerate}
\renewcommand{\labelenumi}{(\roman{enumi})}
\setlength{\itemsep}{1pt}
\setlength{\parskip}{0pt}
\setlength{\parsep}{0pt}
\item $\varphi[x_1,y_1]=0$,
\item for every $j$, $2\le j\le\ell$, $\varphi([x_1,y_j])=\varphi([x_1,y_{j-1}])+1\pmod 3$
if $(y_{j-1},y_j)\in E(\vv{P_\ell})$, and $\varphi([x_1,y_j])=\varphi([x_1,y_{j-1}])-1\pmod 3$
otherwise.
\item for every $i$, $2\le i\le k$, and for every $j$, $1\le j\le\ell$,
$\varphi([x_i,y_j])=\varphi([x_{i-1},y_j])+1\pmod 3$ if $(x_{i-1},x_i)\in E(\vv{P_k})$,
and $\varphi([x_i,y_j])=\varphi([x_{i-1},y_j])-1\pmod 3$ otherwise.
\end{enumerate}
It is then routine to check that the mapping 
$\varphi$ is a homomorphism and thus $\chi_o(\vv{P_k}\cp\vv{P_\ell})\le 3$.
\end{proof}

We now consider
 the upper oriented chromatic number of Cartesian products of undirected
graphs.

\begin{theorem}
If $G$ and $H$ are two undirected graphs with $k=min\{\chi(G),\chi(H)\}$, 
then 
$$\chi_o^+(G\cp H)\le k\chi_o^+(G)\chi_o^+(H).$$
\label{th:cartesian-undirected}
\end{theorem}

\begin{proof}
Assume without loss of generality that $k=\chi(H)$
and let $\lambda$ be a proper $k$-coloring of $H$.
Let $\vv T$ and $\vv U$ be two oriented graphs of order
$\chi_o^+(G)$ and $\chi_o^+(H)$, respectively, such that 
$\vv G\hom\vv T$ for every orientation $\vv G$ of $G$
and $\vv H\hom\vv U$ for every orientation $\vv H$ of $H$.

Let now $\vv{W}$ be the oriented graph defined by
$$V(\vv{W})=\{[\ell,a,b],\ 1\le\ell\le\chi(H),\ a\in V(\vv{T}),\ b\in V(\vv{U})\}$$
and $([\ell,a,b],[\ell',a',b'])\in E(\vv{W})$ if and only if
either $\ell=\ell'$ and $(a,a')\in E(\vv{T})$ or
$\ell\neq\ell'$ and $(b,b')\in E(\vv{U})$.
We shall prove that any orientation of $G\cp H$ admits a homomorphism to $\vv{W}$,
which gives the desired result.

Fix any orientation $\vv{G\cp H}$ of $G\cp H$. 
For every $v\in V(H)$, let $\alpha_v:\vv{G_v}\hom \vv T$ be a homomorphism.
Similarly, for every $u\in V(G)$, let $\beta_u:\vv{H_u}\hom \vv U$ be a homomorphism.

Let $\varphi:V(\vv{G\cp H})\lhom V(\vv{W})$ be the mapping defined
by $\varphi([u,v])=[\lambda(u),\alpha_v(u),\beta_u(v)]$ for every $[u,v]\in V(\vv{G\cp H})$.
We claim that $\varphi$ is a homomorphism. To see this, let $([u,v],[u',v'])$ be an
arc in $\vv{G\cp H}$.
We then have $\varphi([u,v])=[\lambda(u),\alpha_v(u),\beta_u(v)]$
and $\varphi([u',v'])=[\lambda(u'),\alpha_{v'}(u),\beta_{u'}(v)]$.
If $u=u'$ and $(v,v')\in E(\vv{H})$, then $\lambda(u)=\lambda(u')$,
$\beta_u=\beta_{u'}$ and, therefore,
$([\lambda(u),\alpha_v(u),\beta_u(v)],[\lambda(u'),\alpha_{v'}(u),\beta_{u'}(v)])\in E(\vv{W})$.
Similarly, if $v=v'$ and $(u,u')\in E(\vv{G})$, then $\lambda(u)\neq\lambda(u')$,
$\alpha_v=\alpha_{v'}$ and, therefore,
$([\lambda(u),\alpha_v(u),\beta_u(v)],[\lambda(u'),\alpha_{v'}(u),\beta_{u'}(v)])\in E(\vv{W})$.

Every arc in $\vv{G\cp H}$ 
is thus mapped
to an arc in $\vv{W}$ and $\varphi$ is a homomorphism.
\end{proof}

Since $\chi(P)\le 2$ and $\chi_o^+(P)\le 3$ for every path $P$, Theorem~\ref{th:cartesian-undirected} 
implies that $\chi_o^+(P_k\cp P_\ell)\le 2.3.3=18$ for every $k,\ell\ge 3$.
In~\cite{FRR}, Fertin, Raspaud and Roychowdhury (implicitely) proved that the upper oriented
chromatic number of the Cartesian product of any two paths is at most 11.

Moreover, since $\chi(T)\le 2$ and $\chi_o^+(T)\le 3$ for every tree $T$,
we get the following:

\begin{corollary}
Let $T$ be a tree. For every undirected graph $G$, $\chi_o^+(G\cp T)\le 6\chi_o^+(G)$.
\end{corollary}

In the same vein, since the cycle $C_5$ has upper oriented chromatic number 5
and every cycle except $C_5$ has upper oriented chromatic number 4, we obtain:

\begin{corollary}
Let $C_k$ be the cycle on $k$ vertices. For every undirected graph $G$, $\chi_o^+(G\cp C_5)\le 15\chi_o^+(G)$,
and $\chi_o^+(G\cp C_k)\le 12\chi_o^+(G)$ for every $k\ge 3$, $k\neq 5$.
\end{corollary}

These two corollaries improve the results of Natarajan, Narayanan and Subramanian~\cite{NNS},
who proved that $\chi_o(G\cp P_k)\le (2k-1)\chi_o(G)$ and
$\chi_o(G\cp C_k)\le 2k\chi_o(G)$ for every graph $G$.

\section{Direct products}
\label{s:direct}

Since $\vv G\ddp\vv H\subseteq\vv G[\vv H]$ for every two oriented
graphs $\vv G$ and $\vv H$, we get from Theorem~\ref{th:lexicographic-oriented}
that $\chi_o(\vv{G}\ddp\vv{H})\le\chi_o(\vv G)\chi_o(\vv H)$.
This bound can be improved as follows:

\begin{theorem}
For every oriented graphs $\vv{G}$ and $\vv{H}$,
$$\chi_o(\vv{G}\ddp\vv{H})\le\min\{\chi_o(\vv G),\chi_o(\vv H)\}.$$
\label{th:direct-oriented}
\end{theorem}

\begin{proof}
This result directly follows from the fact that
$\vv{G}\ddp\vv{H}\hom\vv G$ and $\vv{G}\ddp\vv{H}\hom\vv H$
since the
mapping $\gamma:V(\vv{G}\ddp\vv{H})\lhom V(\vv G)$ 
(resp. $\gamma:V(\vv{G}\ddp\vv{H})\lhom V(\vv H)$) defined
by $\gamma([u,v])=u$ (resp. $\gamma([u,v])=v$) is clearly a homomorphism
from $\vv{G}\ddp\vv{H}$ to $\vv G$ (resp. to $\vv H$).
\efface{
Assume without loss of generality that $k=\chi_o(\vv G)\le\chi_o(\vv H)$
and let $c$ be an oriented $k$-coloring of $\vv G$.
We claim that the mapping $\gamma:V(\vv{G}\ddp\vv{H})\lhom\{1,2,\dots,k\}$ defined
by $\gamma([u,v])=c(u)$ for every vertex $[u,v]$ is an oriented $k$-coloring
of $\vv{G}\ddp\vv{H}$.
To see this, let $([u,v],[w,x])$ be any arc in $E(\vv{G}\ddp\vv{H})$.
Since $(u,w)$ is an arc in $E(\vv G)$, $\gamma([u,v])=c(u)\neq c(w)=\gamma([w,x])$.
Moreover, if $([u',v'],[w',x'])$ is another arc in $E(\vv{G}\ddp\vv{H})$
such that $\gamma([u,v])=\gamma([w',x'])$, and thus $c(u)=c(x')$,
then $\gamma([w,x])=c(w)\neq c(u')=\gamma([u',v'])$ since $c$ is an oriented coloring.
} 
\end{proof}

We now consider the upper oriented chromatic number of direct products of
undirected graphs.

\begin{theorem}
For every undirected graphs $G$ and $H$ with $k=\chi(G^2)$ and $\ell=\chi(H^2)$,
$$\chi_o^+(G\ddp H)\le \frac{2^{k(\ell-1)}-1}{2^{\ell-1}-1}.$$
\label{th:direct-upper-undirected}
\end{theorem}

\begin{proof}
Let $\vv W$ be the digraph defined by
$$V(\vv W)=\big\{\ [\alpha,\beta,c_{1,1},\dots,c_{1,\ell},\dots,c_{\alpha-1,1},\dots,c_{\alpha-1,\ell}],\ 1\le\alpha\le k,\ 1\le\beta\le\ell,$$
$$c_{i,j}\in\{0,1\},\ c_{i,\beta}=0,\ 1\le i\le\alpha-1,\ 1\le j\le\ell\ \big\}$$
and 
$$([\alpha,\beta,c_{1,1},\dots,c_{1,\ell},\dots,c_{\alpha-1,1},\dots,c_{\alpha-1,\ell}],
[\alpha',\beta',c'_{1,1},\dots,c'_{1,\ell},\dots,c'_{\alpha'-1,1},\dots,c'_{\alpha'-1,\ell}])$$
is an arc in $E(\vv W)$ if and only if one of the following holds:

\begin{enumerate}
\renewcommand{\labelenumi}{(\roman{enumi})}
\setlength{\itemsep}{1pt}
\setlength{\parskip}{0pt}
\setlength{\parsep}{0pt}
\item $\alpha<\alpha'$, $\beta<\beta'$ and $c'_{\alpha,\beta}=1$,
\item $\alpha<\alpha'$, $\beta>\beta'$ and $c'_{\alpha,\beta}=0$,
\item $\alpha>\alpha'$, $\beta<\beta'$ and $c_{\alpha',\beta}=1$,
\item $\alpha>\alpha'$, $\beta>\beta'$ and $c_{\alpha',\beta}=0$.
\end{enumerate}

The digraph $\vv W$ is clearly an oriented graph of order
$1+2^{\ell-1}+\dots+2^{(k-1)(\ell-1)}=\frac{2^{k(\ell-1)}-1}{2^{\ell-1}-1}$.

Let now $\vv{G\ddp H}$ be any orientation of $G\ddp H$,
$a$ be any proper $k$-coloring of $G^2$ and $b$ be any proper $\ell$-coloring of $H^2$.
We shall now construct a mapping
$\varphi$ from $V(\vv{G\ddp H})$ to $V(\vv W)$ and prove
that this mapping
is a homomorphism, which will give the desired result.

Let $[u,v]$ be any vertex in $V(\vv{G\ddp H})$ and 
$$\varphi([u,v])=[\alpha,\beta,c_{1,1},\dots,c_{1,\ell},\dots,c_{\alpha-1,1},\dots,c_{\alpha-1,\ell}]$$ 
be defined as follows:
\begin{enumerate}
\renewcommand{\labelenumi}{(\roman{enumi})}
\setlength{\itemsep}{1pt}
\setlength{\parskip}{0pt}
\setlength{\parsep}{0pt}
\item $\alpha=a(u)$,
\item $\beta=b(v)$,
\item if there is an arc $([u,v],[w,x])$ in $E(\vv{G\ddp H})$ such that 
$a(u)>a(w)$ and $b(v)<b(x)$, then $c_{a(w),b(x)}=1$,
\item if there is an arc $([u,v],[w,x])$ in $E(\vv{G\ddp H})$ such that 
$a(u)>a(w)$ and $b(v)>b(x)$, then $c_{a(w),b(x)}=0$,
\item if there is an arc $([w,x],[u,v])$ in $E(\vv{G\ddp H})$ such that 
$a(u)>a(w)$ and $b(v)<b(x)$, then $c_{a(w),b(x)}=0$,
\item if there is an arc $([w,x],[u,v])$ in $E(\vv{G\ddp H})$ such that 
$a(u)>a(w)$ and $b(v)>b(x)$, then $c_{a(w),b(x)}=1$,
\item every $c_i$ that has not been set in (iii), (iv), (v) or (vi) is set to 0.
\end{enumerate}

Note that in items (iii) to (vi) above, if such an arc exists, then it must be unique since
$a$ and $b$ are proper colorings of $G^2$ and $H^2$, respectively.

Let now $([u,v],[u',v'])$ be any arc in $E(\vv{G\ddp H})$, with
$$\varphi([u,v])=[\alpha,\beta,c_{1,1},\dots,c_{1,\ell},\dots,c_{\alpha-1,1},\dots,c_{\alpha-1,\ell}]$$
and 
$$\varphi([u',v'])=[\alpha',\beta',c'_{1,1},\dots,c'_{1,\ell},\dots,c'_{\alpha'-1,1},\dots,c'_{\alpha'-1,\ell}].$$
We thus have $\{u,u'\}\in E(G)$ and $\{v,v'\}\in E(H)$.

Suppose first that $\alpha=a(u)<a'(u)=\alpha'$.
In that case, if $\beta=b(v)<\beta'=b(v')$ then $c'_{\alpha,b(v)}=1$, otherwise $c'_{\alpha,b(v')}=0$.
Suppose now that $\alpha'=a(u')<a(u)=\alpha$.
In that case, if $\beta=b(v)<\beta'=b(v')$ then $c_{\alpha',b(v)}=1$, otherwise $c_{\alpha',b(v')}=0$.
Therefore, in all cases, $\varphi$ maps the arc $([u,v],[u',v'])$ to some
arc in $E(\vv W)$ and is thus a homomorphism.
\end{proof}

Since $\chi(P^2)=3$ for every path $P$, we get that the direct product
of any two paths has upper oriented chromatic number at most $\frac{2^{3.2}-1}{2^2-1}=\frac{63}{3}=21$.
However, every such product is the disjoint union of two components that are almost grid graphs
and it has been (implicitely) shown in~\cite{FRR} that the 
upper oriented chromatic number of any such component is at
most 11.

\section{Discussion}
\label{s:discussion}

In this paper, we introduced and initiated the study of a new parameter called the upper
oriented chromatic number of undirected graphs which, from our point of view, arises ``naturally''
in the framework of oriented colorings.
We gave some general upper bounds on the upper oriented chromatic number of several types
of product graphs and derived some upper bounds on the ordinary oriented chromatic number
of such graphs.

We hope that this new parameter will motivate further research. In particular, most of our
upper bounds for product graphs
can be significantly improved when considering specific graph classes (a first
step in this direction was our results on products of paths). For instance, it would be
interesting to determine the upper and ordinary oriented chromatic numbers of products
of trees, outerplanar graphs or, more generally, of partial $k$-trees.

In~\cite{SURVEY}, we conjectured that the oriented chromatic number of every {\em connected\,}
cubic graph is at most 7 and, to our knowledge, this conjecture is still open. A related question
is thus the following:

\begin{question}
Determine the upper oriented chromatic number of cubic graphs.
\label{q:upper_cubic}
\end{question}

It has been proved in~\cite{SV} that every orientation of any cubic graph admits a homomorphism
to the Paley tournament of order 11 and, therefore, $\chi_o^+(G)\le 11$ for every cubic graph $G$.
This upper bound can probably be improved. On the other hand, we know that there exists cubic
graphs with oriented chromatic number 7~\cite{SURVEY}.

Every tree has upper oriented chromatic number at most 3 and we know that there are trees with
oriented chromatic number 3. Every such tree $T$ thus satisfies the equality $\chi_o^+(T)=\chi_o(T)$.
This property also holds for every outerplanar graph (or, more generally, for every partial 2-tree)
with oriented chromatic number 7. These observations lead to the following question:

\begin{question}
Give necessary or sufficient conditions for a graph $G$ to
satisfy the equality $\chi_o^+(G)=\chi_o(G)$.
\label{q:chio-chioplus}
\end{question}

The notion of the {\em oriented chromatic index\,} of an oriented graph $\vv G$ has been introduced in~\cite{PS},
and is defined as the smallest order of an oriented graph $\vv T$ such that the {\em line digraph\,}
$LD(\vv G)$ of $\vv G$ admits a homomorphism to $\vv T$.
(Recall that $LD(\vv G)$ is given by $V(LD(\vv G))=E(\vv G)$
and for every two arcs $(u,v)$ and $(u',v')$ in $E(\vv G)$,
$((u,v),(u',v'))$ is an arc in $E(LD(\vv G))$ if and only if $v=u'$.)
In the same vein, we can thus define the {\em upper oriented chromatic index\,} of an undirected
graph $G$ as the
smallest order of an oriented graph $\vv U$ such that the line digraph of every orientation
of $G$ admits a homomorphism to~$\vv U$. Here again, many upper bounds obtained for the
oriented chromatic index of specific graph classes implicitely deal with the upper oriented
chromatic index of these classes. We thus propose to study this other parameter as well.

\end{document}